\documentclass[conference]{IEEEtran}

\usepackage{graphicx}
\usepackage{booktabs}
\usepackage{caption}
\usepackage{subcaption}
\usepackage{amsmath}
\usepackage{amssymb}
\usepackage{algorithm}
\usepackage{algpseudocode}
\usepackage{bbm}

\usepackage{xcolor}
\usepackage{makecell}
\usepackage{url}
\usepackage{comment}
\usepackage{siunitx}
\usepackage{listings}
\usepackage{tabularx}
\usepackage{breqn}

\usepackage{hyperref}
\hypersetup{colorlinks,allcolors=black}

\usepackage{cleveref}
\crefname{figure}{Fig.}{Figs.}
\crefname{section}{Sec.}{Secs.}
\crefname{table}{Tbl.}{Tbls.}
\crefname{equation}{Eq.}{Eqs.}
\crefname{algorithm}{Alg.}{Algs.}
\crefname{appendix}{Appendix}{Appendices}

\usepackage[colorinlistoftodos,textsize=tiny,textwidth=10mm]{todonotes}

\begin{document}
\title{AutoNet: Automatic Reachability Policy Management in Public Cloud Networks}

\author{
{\rm German Sviridov} \qquad {\rm Zheng Tao Shen} \qquad {\rm Jorge Cardoso}\\
Huawei Technologies
}

\maketitle

\begin{abstract} 
Virtual Private Cloud (VPC) is the main network abstraction technology used in public cloud systems. VPCs are composed of a set of network services that permit the definition of complex network reachability properties among internal and external cloud entities such as tenants' VMs or some generic internet nodes. Although hiding the underlying complexity through a comprehensible abstraction layer, manually enforcing particular reachability intents in VPC networks is still notably error-prone and complex.

In this paper, we propose AutoNet, a new model for assisting cloud tenants in managing reachability-based policies in VPC networks. AutoNet is capable of safely generating incremental VPC configurations while satisfying some metric-based high-level intent defined by the tenants. To achieve this goal, we leverage a MaxSAT-based encoding of the network configuration combined with several optimizations to scale to topologies with thousands of nodes. Our results show that the developed system is capable of achieving a sub-second response time for production VPC deployments while still providing fine-grained control over the generated configurations.
\end{abstract}

\section{Introduction}
Public cloud adoption has been observing exponential growth in recent years~\cite{gartner_growth}. Thousands of organizations are either migrating their existing on-premise IT deployments to the cloud or are building their new solutions to be directly cloud-native~\cite{kratzke2017understanding}. 
As more diverse businesses started to move to the cloud, cloud providers started to provide more complex managed services in an attempt to satisfy the particular requirements of their ever-growing customer base. While many major cloud players started with a handful of general-purpose computing and storage services, today these services are in the hundreds and tailored for any possible use case a tenant may have. 
At the same time, most of services related to the management and operation of the fundamental cloud functions remained largely unchanged. Such is the case with cloud network connectivity management. Indeed, cloud networking still employs legacy network architectures composed of subnets, firewalls, routers, and other as-a-service counterparts of traditional network devices.

With the introduction of additional cloud-specific network services, operating cloud networks has reached a complexity high enough to require specialized roles within the enterprises just for such a task. Nevertheless, no matter the amount of available tools and expertise, most of the cloud network configuration changes still remain highly error-prone due to the tangled and complex nature of the cloud network services. Under some conditions, these configuration errors may potentially lead to catastrophic results for the enterprises. Indeed accidentally blocking some services or exposing sensitive information to the public network may result in multi-million losses in revenues~\cite{titania_report} or in hefty fines in the case of severe data breaches~\cite{hoofnagle2019european, solove2020information}.

Major cloud providers have been trying to address issues related to accidental network misconfigurations by offering network reachability debugging tools~\cite{backes2019reachability, azure_net_watcher, aws_vpc_analyzer} to their customers. While these tools are capable of assisting network engineers in finding potential network bugs they do so only after they have been introduced, as opposed to \emph{proactively preventing and mitigating them} through controllable network configuration management. Thus, not providing tenants a way of reliably and controllably deploying network configuration updates within their cloud deployments.

In this paper, we tackle the requirement of proactive cloud network configuration management by proposing AutoNet. AutoNet is designed to assist cloud network engineers by providing a single network-wide abstraction for reachability policy management at the cloud tenants' level. The main goal of AutoNet is that of assisting tenants with incremental network updates by introducing the minimum set of configuration changes with respect to the existing tenant-defined configurations.
To achieve this goal, AutoNet addresses four main requirements that are necessary to guarantee both the system's correctness and its practical usefulness in the public cloud: 
\begin{itemize}
	\item A simple abstraction model to define and manage reachability policies
	\item A second-level response time capable of supporting multiple configuration updates per minute
	\item Formal \emph{network-wide} correctness guarantees of the network configuration changes
	\item Possibility of seamlessly tailoring configuration changes to meet specific high-level objectives such as deployment cost or service performance
\end{itemize}
To achieve these goals, AutoNet employs SAT-based~\cite{ganai2007sat} encoding of the entire cloud network. Additionally, to scale to large topologies, AutoNet makes use of different network compression techniques based on Binary Decision Diagram (BDD)~\cite{akers1978binary}. Finally, AutoNet uses a weighted MaxSAT~\cite{ansotegui2013sat} formulation to allow the system to generate different network configurations based on user-specified functional intents and additional objectives such as configuration complexity, monetary cost, or service performance.

The rest of the paper is organized as follows: in~\cref{sec:background} we describe key concepts behind public cloud networking infrastructure and motivate the importance of automatic reachability policy management. We then present the overall architecture of AutoNet ands describe its core components in~\cref{sec:system-design}. We evaluate our system in~\cref{sec:evaluation} and describe how it compares to existing work in~\cref{sec:related}. Finally, we discuss limitations of our proposal and possible future work in~\cref{sec:discussion}, and draw our conclusions in~\cref{sec:conclusion}.

\section{Public cloud networking}\label{sec:background}
Modern public cloud deployments are composed of a broad set of managed services such as storage, computing, and networking services. Nowadays there exist multiple different flavors of storage and elastic computing services (ECSs) that keep evolving based on the customers' requirements.
Despite the vast service offering, management simplicity remains among the main objectives cloud providers have when developing new managed services for public customers.
Indeed, users are typically exposed to a simplified interface that permits them to easily request new resources and seamlessly interconnect them.
Such a level of abstraction hides most of the underlying complexity away from the users while providing only basic building blocks for building complex and intricate network interconnects for their applications. 

\begin{figure}[t] 		
	\centering		
	\includegraphics[width=0.8\columnwidth]{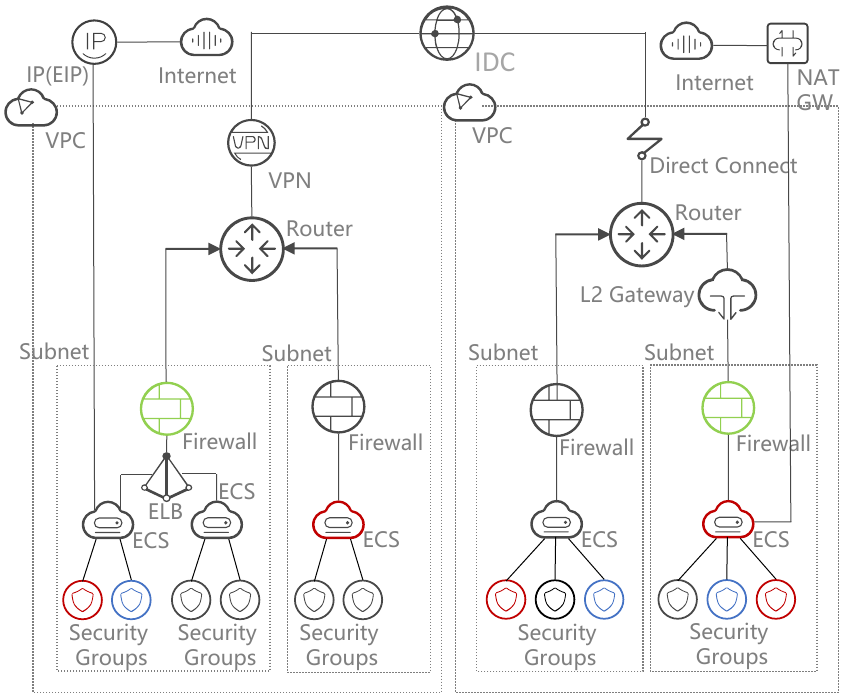}
	\caption{Example of the most commong VPC services and their interconnect}
	\label{fig:vpc_example}
\end{figure}

When it comes to cloud networking, the majority of intra-cloud connectivity services are offered within the Virtual Private Cloud (VPC) macro-service. VPCs include the virtualized counterparts of the traditional network devices such as routers or firewalls, as depicted in \cref{fig:vpc_example}. Their main role is that of providing connectivity between different ECSs, as well as, providing access control, load balancing, and other traffic and security-related features.

\subsection{Virtual Private Cloud architecture}
The main difference that separates VPCs from normal networks is that each service can be located only at designated locations within the network topology. To do so, VPCs are built in a structured and hierarchical way as depicted in \cref{fig:vpc_example}. 
ECSs are aggregated into logical groups within a VPC using virtual subnets and are connected through one or more virtual NIC. Finally, multiple subnets are grouped into logical VPCs with an N:1 mapping among them.
In addition to traditional services, VPCs may include direct connect network gateways which may provide load balancing through elastic load balancers (ELBs), connectivity to external networks, to other VPCs within the cloud infrastructure, or to the internet network through Elastic IPs (EIPs) or NAT gateways (NAT-GWs).

To provide traffic control over the traffic flows, VPC services are typically associated with one or more routing or filtering service. These services are configured with a corresponding configuration that defines, e.g., a routing table for routers, a remote destination for VPC Peerings, a public IP address for EIP, or ACL rules for network ACLs (NACLs) and security groups (SGs), etc. 
The entirety of the configurations of all of the VPC network services defines the overall VPC \emph{network reachability policy}. A reachability policy is a mapping between all pairs of VPC services representing the traffic source and the traffic destination and a corresponding header space representing the allowed traffic between the two VPC services. 
Although limited, the exposed set of configurations within the VPC macro-service can still lead to an explosion in complexity when customers' topologies grow larger, in the presence of non-out-of-the-box deployments (e.g., hybrid or multi-cloud deployments), or simply when the existing network services are not properly managed.
Indeed, differently from the mapping of VPC services, individual configurations can be assigned to multiple service instances. This relation implies the possibility of the following scenarios manifesting:
\begin{itemize}
	\item \textbf{Replicated configurations:} a single configuration can be associated with multiple distinct network services. This implies that, e.g., the same set of ACL rules can reside within two different NACLs. Thus changing the configuration of one network service may affect the configuration of another network service. \cref{fig:vpc_example} shows a possible instance of this behavior with the first and last subnets sharing a green firewall.
	\item \textbf{Tangled configurations:} multiple network services can have more than one associated configuration which are treated as a single configuration according to some configuration merging mechanism. E.g., a single router can be associated with a list of different routing tables, or a single subnet can be associated with multiple NACLs that are applied to the transient traffic according to their priority. \cref{fig:vpc_example} depicts such possibility with first and last ECSs sharing a red and blue SGs.
	\item \textbf{Replicated network addresses:} two or more VPC network services can share the same network address space. In \cref{fig:vpc_example} a possible example of this scenario could be the two red ECSs, with one being located on an on-premise deployment and the other being deployed inside the cloud data center.
\end{itemize}

The aforementioned three properties imply that \emph{a change to a potentially localized configuration can have a broader effect on the entire network} if the network configuration is not managed properly. This, in turn, makes it even more difficult to verify or enforce desired reachability intents without introducing misconfigurations and potentially compromising the correctness of the entire network.

\subsection{Reachability policy management in VPC networks}

Most cloud customers nowadays spend considerable amount of resources towards continuous assurance of reachability policy intents within networks. Such reachability policies may include some intents related to the security of the services, such as access to some sensitive storage from the public networks, or to some functional intents such as guaranteeing connectivity between the front-end services and the back-end databases. In \cref{tbl:intents} we summarize some of the most common reachability policy intents that customers typically want to implement within their deployments.

\subsubsection{The difficulty of verifying reachability properties:} 
Most of the reachability policy intent assurance systems still employ traditional network probing~\cite{guo2015pingmesh} capable of only providing weak guarantees about the network behavior. While such approaches can lead to meaningful coverage in small deployments, due to the ever-growing complexity and scale of modern VPCs, continuously and provably assuring full test coverage over the entire network keeps becoming more and more challenging~\cite{xu2021test}.
To tackle this problem, major cloud providers have been offering VPC reachability verification tools~\cite{backes2019reachability, azure_net_watcher, batfish, aws_vpc_analyzer} capable of verifying whether a given reachability intent is satisfied or not within the tenants' deployment. Yet, while addressing the original problem, whenever a misconfiguration or a policy violation is found, customers are still required to manually act to fix the problem. This may lead to an iterative and tedious trial and error approach which may require up to hours of manual labor for large topologies~\cite{potharaju2013network} with complex intents~\footnote{Estimation based on the measured average response time of commercial network verification services}.
When considering that cloud network configurations can be operated on up to several times per hour, even a small degree of misconfiguration probability may lead to the emergence of severe bugs~\cite{kumar2019cloud} that are notoriously difficult to detect and may persist over a prolonged amount of time. 

\subsubsection{The necessity of satisfying high-level objectives:}
While a network configuration can be functionally correct by itself and satisfy all functional requirements in terms of desired reachability properties, it may still not satisfy some high-level cost or performance metric. 
Due to the presence of multiple different network services, in most of the existing public cloud providers, there may be multiple alternatives to satisfy the same functional objective. Although being functionally equivalent, each alternative can lead to different properties in terms cost, security, or performance. Indeed, connecting the front-end server to the in-house database through a direct connect would lead to the highest possible service performance and security while incurring a substantial cost for the business. On the other hand, letting the traffic transit through the public internet instead will lead to the opposite effect. The choice among these alternatives, alongside to functional requirements, is usually left to the cloud network engineer to satisfy.
In such a complex setting guaranteeing that a particular change is both correct and optimal in terms of some metric becomes virtually impossible without specialized tools.
To the best of our knowledge, no cloud provider has been able to provide an effective system capable of jointly enforcing and verifying a given functional requirement while satisfying additional high-level metrics in a fully automated and transparent way.

\begin{table}[t]
	\small
	\centering
	\caption{Example reachability policy intents to be enforced in a commercial cloud deployment}
	\label{tbl:intents}
	\renewcommand{\arraystretch}{1.3}
	\begin{tabularx}{\linewidth}{lX}
		\toprule
		\textbf{\makecell{Intent}} & \textbf{\makecell{Intent description}} \\ \midrule
		\textbf{I1 - Simple Reachability} & Some traffic type is allowed or denied between two ECSs \\
		\textbf{I2 - Internet Isolation} & Some traffic type is allowed or denied between an ECS and some internet node \\
		\textbf{I3 - Subnet Isolation} & Some traffic type is allowed or denied between two subnets \\
		\textbf{I4 - Subnet Internet Isolation} & Some traffic type is allowed or denied between a subnet and some internet node \\
		\textbf{I5 - Waypointing} & Force some 5-tuple traffic between two ECSs to traverse a third ECS \\
    \bottomrule
	\end{tabularx}
\end{table}

\section{AutoNet: Automatic reachability policy management for VPC networks}\label{sec:system-design} 
AutoNet aims at providing a scalable cloud network management tool capable of transparently enforcing new reachability policies at a second-level scale for large VPC deployments. At the same time, AutoNet aims at preserving the properties of the existing reachability policies while also satisfying high-level configuration objectives defined by the operators. 

The properties AutoNet tries to achieve have been shown to be an EXPSPACE-complete and NP-complete problem in the general case~\cite{schneider2022complexity, velner2016some}. To tackle this complexity and to achieve our functional and performance goals we employ a set of suitably tailored techniques that permit to reduce the dimensionality of the problem, and at the same time safely and efficiently generate updated reachability policies based on user-defined intents. Most notably, AutoNet is based on the following core components that are later discussed in detail: 
\begin{itemize}
	\item \textbf{Definition of reachability intent (\cref{sec:intent_definition})}: Provides a way for the user to define a reachability policy intent among the one present in ~\cref{tbl:intents} as well as configuration weights used to steer AutoNet towards more desirable solutions
	\item \textbf{Network topology abstraction (\cref{sec:topology_abstraction})}: Provides a common abstraction for all of the complex network services present in the cloud deployments and permits to reduce the entire VPC topology to a simple to manage tree-based topology
	\item \textbf{Network pre-processing (\cref{sec:net_preprocessing})}: Allows to reduce the dimensionality of the problem by pruning part of the services present in the cloud deployment and/or by compressing their configurations 
	\item \textbf{SAT-based actionable network reachability encoding (\cref{sec:sat_encoding})}: Allows to formally encode reachability properties among different services in the cloud deployment by reasoning about the topological properties of the VPC deployment as opposed to path-based based techniques
	\item \textbf{MaxSAT-based configuration update synthesis (\cref{sec:maxsat})}: Allows to compute the minimum set of configurations to be changed to enforce some particular reachability policy while satisfying users' intents in terms of desired solution space
\end{itemize}

\subsection{Definition of reachability intent}\label{sec:intent_definition}
Similarly to the existing tools~\cite{backes2019reachability, azure_net_watcher}, AutoNet requires users to specify the desired reachability intent. In AutoNet we define an intent as a particular desired reachability policy composed of i) a desired traffic action such as allow or deny, ii) an intent header space (IHS) which represents the header space for which the intent must be satisfied, iii) a set of source and destination nodes between which the intent must hold, and, optionally, iv) a set of configuration weights that permit the user to steer AutoNet towards the generation of some specific configurations (e.g., preferring the changes to the configuration of existing network services as opposed to the generation of new ones).

\subsection{Network topology abstraction}\label{sec:topology_abstraction}
Cloud deployments may include a multitude of different complex services that provide networking capabilities. 
Nevertheless, it can be noticed that almost all of these services can be reduced to a small set of abstract services. Indeed, these services can be broadly classified into three main abstract network services:
\begin{itemize}
	\item \textbf{Filtering service}: defined as network services capable of allowing or denying (i.e., filtering) at least a part of the traffic that traverses them (e.g., NACLs, SGs, firewalls). The associated configuration is composed of a list of rules that map some header space to an allow/deny action.
	\item \textbf{Routing service}: defined as network services capable of altering the path of at least a part of the traffic that traverses them (e.g., routers, ELBs, transit gateways, etc.). The associated configuration is composed of a list of rules which map some header space to a next-hop network service.
	\item \textbf{End devices}: defined as network services that act as one of the endpoints for a traffic flow. Such services can be either a source or a destination of a particular traffic flow (e.g., ECSs, BMSs). The associated configuration is typically composed of at least an associated unique ID, but in most cases includes also an associated IP address.
\end{itemize}

We noticed that this set of basic primitive services is sufficient to model most of the network services present in a cloud deployment and we use this abstraction in AutoNet to simplify the modeling and analysis of the multitude of network services present in commercial cloud providers by representing each of them as a combination of the three primitive classes. 

Additionally, by observing \cref{fig:vpc_example}, it can be seen that, apart from a handful of network services, most of the services within a VPC are connected in a tree topology. To achieve a full tree topology move the out-of-place network services to an ad-hoc created VPC while also installing an equivalent set of immutable rules on all of the intermediate nodes to provide reachability equivalence to the original topology.

\subsection{Network pre-processing}\label{sec:net_preprocessing}
While the network topology abstraction permits us to make the problem more approachable, it does not help in reducing the algorithmic complexity. 
To tackle this problem we employ two network pre-processing techniques, namely \emph{topology pruning}, and \emph{configuration quantization}. The main goal of these techniques is to filter out and remove some of the abstract services from the network topology or to at least simplify their configuration.
\subsubsection{Topology pruning} aims at reducing the total amount of network services present in the final model of the network. For this purpose, all of the endpoints which do not lead to source or destination network address space overlap with the specified IHS are removed from the topology. The rationale behind this processing step is that, even if the IHS traffic between the filtered-out end devices is affected by the configuration change, the overall reachability policy will remain invariant due to the default traffic filtering at end devices. Exceptions to the aforementioned scheme (e.g., EIPs that provide connectivity to the entirety of public IPs) exist and are managed accordingly by not applying any pruning. 

\subsubsection{Configuration quantization} aims at quantifying the behavior of a network service in response to the IHS. For the purpose of quantization, service configurations are translated into their corresponding BDD representation. Similarly to~\cite{batfish}, for both filtering and routing services, the quantization is performed by iteratively applying a sequence of boolean operations on the BDD representation of individual configuration lines.
For network filters, quantization implies querying the associated configuration for the response in case of an incoming/outgoing IHS flow. Such a response can either be an “allow”, a “deny” or a partial result. The latter implies that only a part of the IHS is allowed to flow through the network filter. In such cases, the result is substituted with the opposite of the desired reachability property (e.g., if the desired reachability property is “allow” the partial quantified result is substituted with “deny”). The rationale behind this is that we want the entirety of the IHS defined in the intent to satisfy the desired reachability property. After the quantization has been performed each configuration file associated with network filters is substituted with the quantified filtering result (i.e., “allow” or “deny”). 
For network routers configuration quantization involves extracting the set of possible reachable devices for the IHS defined within the associated configuration. The possible reachable devices are extracted by considering the next hop field within the associated configuration. After the quantization has been performed each configuration file associated with network routers is substituted with the extracted set.

\subsection{Network encoding}\label{sec:sat_encoding}
The network encoding step involves the translation of the quantified and sampled network into an SAT-based representation of the entire network reachability policy. The resulting encoding will be used later by our MaxSAT to enforce reachability intents.
In the following we provide a simplified formulation of the aforementioned constraints using the developed model and discuss the scalability of the proposed approach. For the purpose of the formulation, we assume to have an unquantified and un-quantized 3-layer tree topology. We also assume that all nodes have a single configuration as opposed to having a distinct set of associated ingress and egress configurations.

All configurations $f \in F$ present in the abstracted network are associated with a set of header spaces that summarize their behavior. Such header spaces are defined in terms of explicitly allowed, denied, and matched header space, namely $c_f^{(A)}$, $c_f^{(D)}$, $c_f^{(M)}$. Since a single node can be associated with multiple configurations we need to define a suitable node configuration encoding function $c(f_1,..,f_n)$ which translates all of the configurations associated with a node to their synthetic representation. Such a function is defined recursively as:

\begin{multline}\label{eq:filter_encoding}
	c(f_1,..,f_n) = c_{f_1}^{(M)} \rightarrow \\ (c_{f_1}^{(A)} \wedge \lnot c_{f_1}^{(D)}) \wedge (\lnot c_{f_1}^{(M)} \rightarrow c(f_2,..,f_n))
\end{multline}

The equation \cref{eq:filter_encoding} encodes the multi-level matching of different configurations assuming a decreasing matching priority (i.e., $f_1$ has the highest priority).
Following this definition, we can define the total admissible header space over a generic node $n \in N$ with of all of the associated configurations $F_n \in F$ as $R_n = c(F_n)$.

Using $P:  N  \rightarrow  N $ as a function that maps each node to the corresponding parent node, the total admissible traffic between two nodes $s,d \in N$ can be expressed as follows:

\begin{multline}\label{eq:all_encoding}
	R(s,d) = (R_s \wedge R_d) \wedge (\lnot (P(s) = P(d)) \rightarrow \\ (R_{P(s)} \wedge R_{P(d)})) \wedge (\lnot (P(P(s)) = P(P(d))) \rightarrow \\ (R_{P(P(s))} \wedge R_{P(P(d))})) \\, \forall s \in  N , \forall d \in  N 
\end{multline}

The key peculiarity of \cref{eq:all_encoding} is that the entire encoding can be performed in $O(N)$. Indeed, \cref{eq:all_encoding} can be easily rewritten as a series of hierarchical clauses, thus not requiring, neither enumerating all of the possible paths a flow can take nor building $O(N^2)$ equations for all of the possible combinations of source and destination pairs.

Our goal is to devise an algorithm capable of altering the above formulation in order to achieve the desired target reachability policy for a particular set of $s,d \in N$. To do so, we need a way of altering the previously formulated expressions to include a set of decision variables. These decision variables allow us to bypass the normal expression for the admissible traffic on any given configuration or network device. For network configurations, this is done by encoding the condition \emph{"if the decision variable is set to some header space then return the set header space, otherwise return the normal admissible header space"}. If we want to support both allow and deny queries we have to introduce two types of decision variables for each network service and each configuration. The new expression becomes a ternary expression of the form \emph{"if the allow decision variable is set to some header space then return the header space, else if it the deny decision variable is set to some header space, return the negated header space, otherwise return the normal admissible header space"}. For this reason, we introduce an \emph{actionable equivalent} expression of \cref{eq:filter_encoding} $a(f_1,..,f_n)$ defined as:

\begin{multline}\label{eq:actionable_filter_encoding}
	a(f_1,..,f_n) = ((\phi_{f_1}^{(A)} \vee \phi_{f_1}^{(D)}) \rightarrow (\phi_{f_1}^{(A)} \wedge \lnot \phi_{f_1}^{(D)})) \vee c_{f_1}^{(M)} \\ \rightarrow (c_{f_1}^{(A)} \wedge \lnot c_{f_1}^{(D)}) \wedge (\lnot c_{f_1}^{(M)} \rightarrow a(f_2,..,f_n))
\end{multline}

$\phi_{f_i}^{(D)}$ and $\phi_{f_i}^{(A)}$ represent respectively the decision variables that, once assigned, permit to respectively add an explicit deny or allow for the IHS within a set of configurations $f_i$ associated with a single node\footnote{For the sake of simplicity we omit the decision variables for the creation of new routing entries}. 

Nevertheless, in some cases, there may exist scenarios that would make the problem unsolvable without adding additional configurations to the network (e.g., missing EIP for internet access). For this reason, we introduce an additional set of node ternary decision variables which encode the possibility of generating new configurations (e.g., NACLs, EIPs, routing tables, etc) from scratch as:

\begin{equation}\label{eq:allow_creation_of_new_elements}
	A_n = ((\psi_n^{(A)} \vee \psi_n^{(D)}) \rightarrow (\psi_n^{(A)} \wedge \lnot \psi_n^{(D)})) \vee a(F_n)
\end{equation}

\cref{eq:allow_creation_of_new_elements} encodes the aforementioned requirement for each node $n$ by permitting the assignment of some values to the decisional variables $\psi_n^{(\cdot)}$ in a similar way to \cref{eq:actionable_filter_encoding}. Given the aforementioned expression, we can define the \emph{actionable configuration expression} $A(s, d, \Phi^{(A)}, \Phi^{(D)}, \Psi^{(A)}, \Psi^{(D)})$ in an equivalent way to \cref{eq:all_encoding}.

Finally, given a desired reachability intent expressed in the form of a set of source and destination nodes $N^{(\texttt{IHS}, \texttt{SRC})}, N^{(\texttt{IHS}, \texttt{DST})}$, and a particular IHS $h^{\texttt{IHS}}$, we can formalize the reachability intent satisfaction and the reachability policy conservation constraints respectively as:

\begin{multline}\label{eq:intent_satisfaction}
	\exists\Phi^{(A)},\ \Phi^{(D)},\ \Psi^{(A)},\ \Psi^{(D)} | \\
	A(s, d, \Phi^{(A)}, \Phi^{(D)}, \Psi^{(A)}, \Psi^{(D)})= R(s, d)\vee h^{\texttt{IHS}} \\,\forall s\in N^{(\texttt{IHS}, \texttt{SRC})},\ \forall d\in N^{(\texttt{IHS}, \texttt{DST})}
\end{multline}

\begin{multline}\label{eq:reachability_conservation}
	\exists \Phi^{(A)}, \Phi^{(D)}, \Psi^{(A)}, Psi^{(D)} | \\
	 A(s, d, \Phi^{(A)}, \Phi^{(D)}, \Psi^{(A)}, \Psi^{(D)})= R(s, d) \\,\forall s \neq d \in  N  \setminus ( N^{(\texttt{IHS}, \texttt{SRC})} \cup  N^{(\texttt{IHS}, \texttt{DST})})
\end{multline}

\cref{eq:intent_satisfaction} guarantees that the desired reachability intent will be enforced while \cref{eq:reachability_conservation} guarantees that at the same time, no accidental changes to the reachability properties of unrelated network elements will be introduced.
Satisfying these constraints requires finding a suitable assignment for all services in $\Phi^{(A)}, \Phi^{(D)}, \Psi^{(A)}, \Psi^{(D)}$. Such an assignment will guarantee that only the reachability policy between $ N^{(\texttt{IHS}, \texttt{SRC})}$ and $ N^{(\texttt{IHS}, \texttt{DST})}$ will be affected while all of the other end-to-end reachability policies will remain unaffected. 

Throughout the formulation, we use only two decision variables per configuration element which is enough to cover all reachability scenarios involving intra-cloud communication. Indeed, a single decision variable can model a single header space and since intra-cloud communication does not involve packet header overwriting each configuration will require at most two rules to be added/changed to enforce a given reachability property (represented by $\phi_{f_i}^{(A)}$, $\phi_{f_i}^{(D)}$). For inter-cloud communication we jointly solve multiple instances of the same problem with different network quantization applied to the network based on the actual header space transformation performed by network services.

\subsection{Configuration generation}\label{sec:maxsat}
To find a suitable solution to the encoded problem we can formalize the problem as a minimization problem. We want to minimize the number of decision variables set to a value different from the default value (i.e., $\emptyset$). We can formulate this requirement as follows:

\begin{equation}\label{eq:maxsat_solution}
\min{\left(\sum\omega_i\left(\phi\right._i\neq\emptyset\mathrm{)\ } \right)}\forall\phi_i\in(\Phi^{\left(A\right)}\cup\Phi^{\left(D\right)}\cup\ \Psi^{\left(A\right)}\cup\ \Psi^{\left(D\right)})
\end{equation}

Solving this objective function will lead to the minimum assignment for all of the decision variables. 
To solve the constrained objective function, we use a MaxSAT formulation and solve it using a weighted implementation of the PmRes algorithm from~\cite{narodytska2014maximum}. The formulation treats AutoNet's decision variables as weighted soft constraints and uses a core-guided approach to find the minimum weighted set of unsatisfiable soft constraints that will satisfy the hard constraints in \cref{eq:intent_satisfaction} and \cref{eq:reachability_conservation}.

\subsubsection{Objective-aware configuration updates}
Weight factors $\omega_i$ permit us to steer the solution algorithm towards desired configurations. This permits to prioritize configurations that, e.g., introduce less explicit deny rules (larger weights for $\phi\in\Phi^{\left(D\right)})$ or avoid blackhole routing rules (larger weights for $\phi_f\in\left(\Phi^{\left(A\right)}\cup\Phi^{\left(D\right)}\right),\ \ \forall f\in P\left(s\right),\ \forall s\in N _e$). At the same time, a suitable assignment to the weight factors can encode high level-objectives such as monetary cost. As an example, one can require to enforce reachability between all hosts within a subnet and an external IP address. By using $\omega_i$ proportional to the hourly cost of an EIP and NAT-GW services AutoNet will be able to automatically find which service is more convenient to deploy under the current scenario. Similarly, one can assign $\omega_i$ based on the average amount of bandwidth multiplied by the price per GB of transferred data flowing between on-premise deployment and the public cloud. AutoNet will be capable of selecting the best solution among, e.g., VPN, EIP, or Direct Connect, to be used to satisfy the requirements while minimizing the price of the intent. Finally, under the same scenario, $\omega_i$ can express the security requirements of a reachability intent. Indeed, high-security requirements may limit the aforementioned choice to VPN and Direct Connect only as they guarantee that the traffic will either be encrypted or will flow exclusively between two secured endpoints.

\subsubsection{Translation to VPC configurations}
The outcome of the MaxSAT algorithm is a suitable assignment to the decision variables. Yet, these decision variables need to be synthesized into actual configurations to be applied to the network services. For this reason, we use a specific configuration translation engine capable of mapping each assignment into a corresponding change in a configuration of a particular network service. The actual deployment of the configuration changes depends on the specific network service. For simple filtering or routing services, an additional allow/deny rule or a routing table entry for the IHS is added in the top position of the affected configurations and eventual anomalies are removed~\cite{khorchani2012firewall}. This technique is guaranteed to lead to a correct final configuration as it mimics exactly the behavior formalized in the actionable node configuration encoding function equation in \cref{eq:actionable_filter_encoding}. In the case of the generation of new network services, such as in the case of a creation of a new EIP or ELB, the procedure will follow a the normal flow for the assignment of the configuration parameters of each service while requiring minimum input from the tenant (e.g., assigning metadata to new services). Finally, the generated configuration change can then be directly applied to the network following standard approaches such as cloud SDK or through the cloud management web console.

\section{Experimental evaluation}\label{sec:evaluation}
In this Section, we describe the implementation details and the evaluation of AutoNet with both synthetic and real data. The take-home messages of our evaluation can be briefly summarized in the following:
\begin{itemize}
	\item AutoNet can perform reachability policy enforcements in under 10 seconds for 2K involved ECSs on top of a commodity machine.
	\item Total running time of AutoNet scales linearly with the number of ECSs involved in the reachability policy enforcement.
	\item The time required to perform SAT encoding of the cloud deployment increases linearly with the number of involved subnets and VPCs.
	\item AutoNet can enforce all of the reachability intents from \cref{tbl:intents} in under 250ms for medium-to-large size cloud deployments
\end{itemize}

\subsection{Implementation}
We implemented AutoNet on top of Batfish~\cite{batfish} as a separate module with 6K lines of Java code.  To be able to parse 
cloud configuration files and translate them to a vendor-independent representation we also extend the Batfish parser with extra logic. Finally, we extend the JavaBDD library used by Batfish to support multithreading and accelerate the network-preprocessing step of AutoNet. To perform the network encoding and generate final solutions we use Z3~\cite{de2008z3} SMT solver version 4.8.12.

We evaluated AutoNet by running the system on top of a 
cloud compute instance equipped with 4 virtual CPUs and 8GB of RAM. The entirety of AutoNet runs as a single-threaded application.
For our evaluation, we consider synthetically generated VPC configurations as well as some snapshots of real configurations of internal customers.
The synthetic topology generator accepts as inputs the number of ECSs, total number of subnets, total number of VPCs, total number of unique SGs, NACLs, and routers, and inter-VPC peering probability. Additionally, to simulate tangled configurations the generator accepts the average number of SGs per ECS. 
The generator will randomly generate VPC topologies by assigning each ECS to a random subnet, and each subnet to a random VPC. A similar procedure is done for filtering and routing services.

\subsection{Scalability analysis}
\begin{figure}[!t]
	\centering
	\begin{subfigure}[b]{0.5\columnwidth}
		\includegraphics[width=1\linewidth]{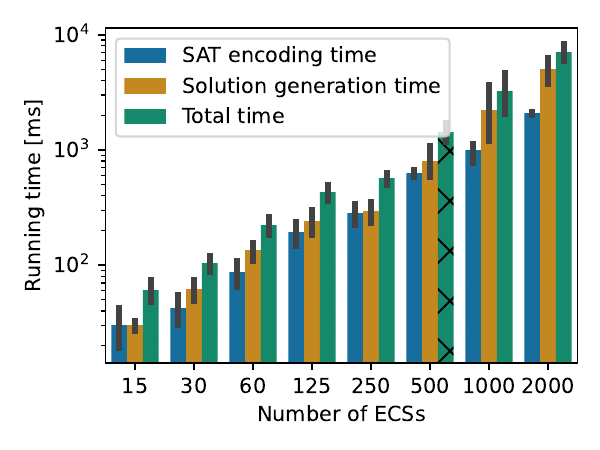}
		\caption{Breakdown of AutoNet running time}
		\label{fig:running_time_random_broad}
	\end{subfigure}%
	~ 
	\begin{subfigure}[b]{0.5\columnwidth}
		\includegraphics[width=1\linewidth]{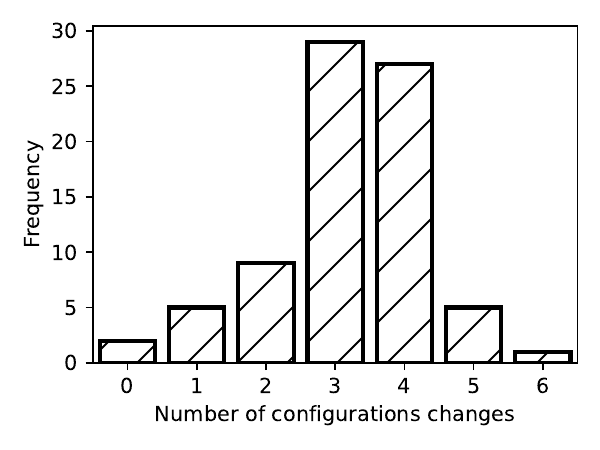}
		\caption{Distribution of the amount of configuration changes}
		\label{fig:configuration_distribution_random_broad}
	\end{subfigure}
	\caption{Scalability of AutoNet for reachability policy enforcement in a random topology for varying number of ECSs}
	\label{fig:random_broad}
\end{figure}

We evaluate the scalability of our system by considering random topologies of a varying number of ECSs. We fix the number of VPCs equal to 4, the number of subnets, NACLs and routers equal to 10, and the average number of SGs per ECS equal to 10. Unless, explicitly stated we use unitary weights for all of the possible decision variables a part from the generation of new network services which are set to a weight of 10.
We consider two random ECSs $e_s, e_d$ within the generated topologies and define the reachability policy to be enforced as \emph{"allow all traffic originating from $e_s$ and destined to any other host to also reach $e_d$"} (i.e., all of the destination fields of the reachability header space are set to wildcard).  Under such reachability intent, the topology pruning will not be able to filter out any node from the topology due to the presence of a wildcard destination header space, thus forcing the SAT encoding to encode the entire network. 
We run all of the experiments 10 times and analyze the breakdown of the running time by considering the time required to generate the SAT encoding of the network and the time to generate the solution in terms of actual configurations to change. \cref{fig:running_time_random_broad} shows that even for topologies with 2000 ECSs the total running time stays below 10 seconds. The majority of the time is spent in trying to find a satisfiable solution with the minimum cost, while the time required to perform SAT encoding of the network stays in the order of magnitude of 1 second. For the same experiments \cref{fig:configuration_distribution_random_broad} depicts the distribution of the number of changes required to enforce the desired reachability policy. Most of the instances require between 3 to 5 unique configuration changes with some of the instances requiring up to 7 unique configuration changes. In all cases, the majority of these configuration changes relate to a simple addition of new ACL or routing line within existing SG/NACLs and routing tables.

\begin{figure}[!t]
	\centering
	\begin{subfigure}[b]{0.5\columnwidth}
		\includegraphics[width=1\linewidth]{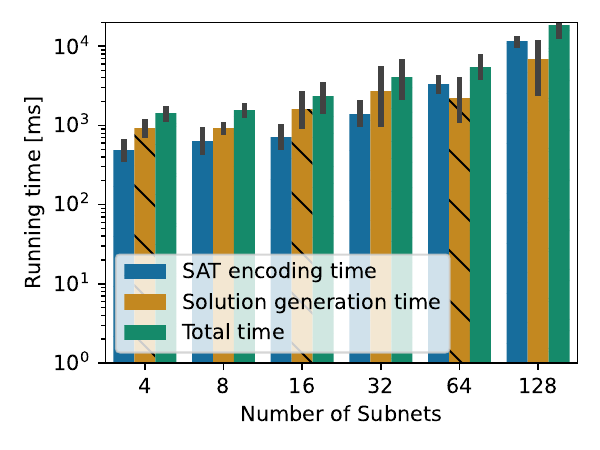}
		\caption{Variable number of subnets}
		\label{fig:running_time_random_broad_subnets}
	\end{subfigure}%
	~ 
	\begin{subfigure}[b]{0.5\columnwidth}
		\includegraphics[width=1\linewidth]{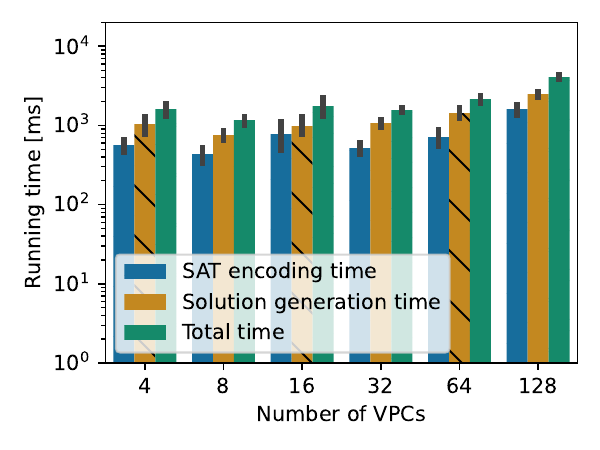}
		\caption{Variable number of VPCs}
		\label{fig:running_time_random_broad_vpcs}
	\end{subfigure}
	\caption{Scalability of AutoNet for reachability policy enforcement in a random topology for varying number of subnets and VPCs}
	\label{fig:random_broad_vpcs_subnets}
\end{figure}

\begin{figure}[t]
	\centering
	\includegraphics[width=1\columnwidth]{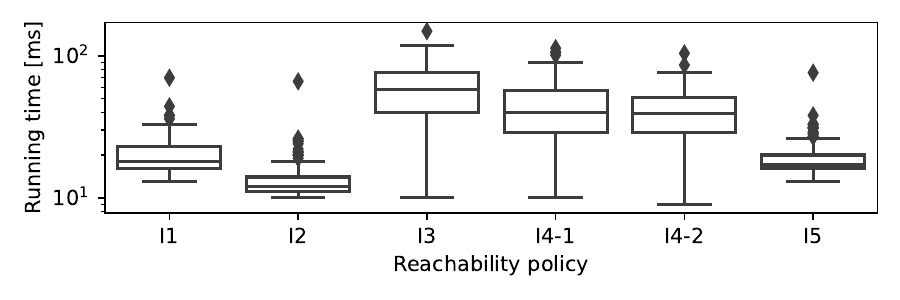}
	\caption{Running times for intents in \cref{tbl:intents} for synthetic topologies. I4-1 and I4-2 represent the exclusive use of, respectively, EIPs and NAT-GW.}
	\label{fig:intents_random}
\end{figure}

Using the same setting from the previous experiments we now fix the total number of ECSs to 500 and assess the scalability of AutoNet in function of the number of subnets and VPCs. \cref{fig:running_time_random_broad_subnets} depicts the running time breakdown for the scenario with a varying number of subnets. As the amount of subnets grows larger AutoNet takes more time to perform the SAT encoding of the network. Notably, for 128 unique subnets the encoding time approaches the 10 seconds mark. At the same time, the total time required to find the solution to the policy enforcement problem does not increase with the number of subnets. This happens due to the fact that each subnet does not contribute individually to the increase in the number of decision variables. Indeed, an increased number of subnets only contributes to the number of expressions of degree 3 generated in \cref{eq:all_encoding}. 
We next performed results to assess the scalability of AutoNet for a variable number of VPCs.
For this purpose we set the number of ECSs to 500 and the number of subnets equal to the number of VPCs. The breakdown of the running times under such a scenario is depicted in \cref{fig:running_time_random_broad_vpcs}. Similarly to the scenario with a variable number of subnets the total running time increases. Yet, under such a scenario, the network encoding contributes less to the total running time. 

For both the experiments with variable VPCs and subnet sizes results similar to the variable ECSs case have been obtained for the distribution of the number of configuration changes and are omitted.

\subsection{Intent performance analysis}\label{sec:intent_evalaution}
We now consider a more realistic analysis in which we use a more strict reachability enforcement intent and we consider more different reachability intents. Notably, we evaluate AutoNet for the intents from \cref{tbl:intents} and report our findings in \cref{fig:intents_random}. We consider both allow and deny counterparts of various intents. Yet, due to the way random topologies are generated most of the flows are denied by default, thus we report only the results for the allow case. \cref{fig:intents_random} shows that for simple allow between two ECSs (I1), the running time stays just above 20ms. Similarly, enforcing reachability between a single ECS and a generic internet node (I2) the total required time stays around 14ms. This behavior is mainly due to the fact that fewer portions of the network need to be analyzed under such a scenario as most of the time this operation only requires a change to a single security group and the creation of a new EIP. The running times increase when it comes to more broader intents such as subnet-to-subnet reachability enforcement (I3). Under such scenario, the average number of actions required is 6.15, as opposed to the I1 and I2 which require 3.69 and 2.11 actions respectively. Additionally, we evaluate the effect of the weights in \cref{eq:maxsat_solution} by considering subnet-to-internet reachability enforcement intents (I4-1, I4-2). For this scenario, we consider two sets of distinct weights. For I4-1 we set the cost of adding a new NAT-GW to infinity, thus forcing the solver to assign individual EIPs to each ECS belonging to a subnet. On the contrary, in I4-2 we perform the opposite and force the solver to only use NAT-GW. The two solutions present a similar performance although the number of required actions varies greatly (8.10, and 5.85 respectively). Finally, we evaluate the performance of AutoNet for the waypoint enforcement intent (I5). Under this scenario the solver behaves similarly to I1, as the overall complexity of the solution is similar,  leading to an average running time of 19.71ms. 

\subsection{Evaluation with a real deployment}
In addition to the synthetic VPC topologies, we were able to evaluate AutoNet on top of a medium-size real cloud deployment that is used by an internal customer. The deployment includes applications related to big data processing, rendering and front/back-end web services running on top of 300-500 ECSs, 30-50 VPCs 10+ NAT-GW services, 10+ ELBs, and 50+ EIPs. Most of the access control is managed by SGs with 80+ distinct SG configurations, with the presence of replicated network addresses, as well as, a medium degree of replicated and tangled configurations. 

\begin{figure}[t]
	\centering
	\includegraphics[width=1\columnwidth]{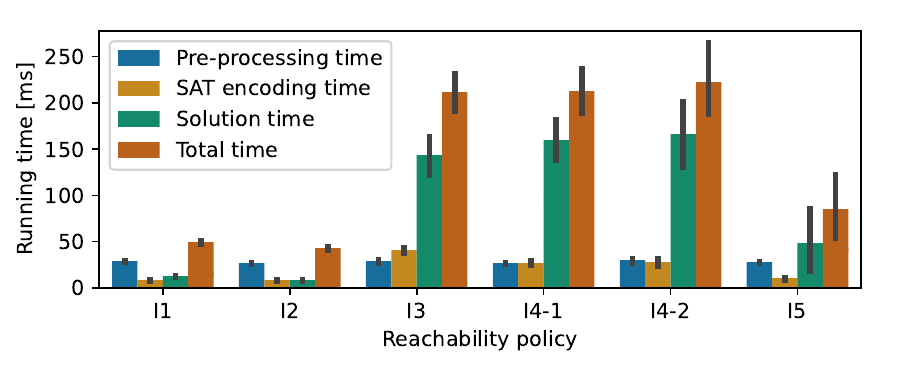}
	\caption{Running times for intents in \cref{tbl:intents} for a real VPC deployment.}
	\label{fig:intents_cn5}
\end{figure}

We perform the same experiments performed in \cref{sec:intent_evalaution} while additionally plotting also the time required to perform the network pre-processing (i.e., quantization and filtering). We consider a random sampling of source and destination nodes for all intents and repeat each run 1000 times. \cref{fig:intents_cn5} shows the results in terms of running time for all of the considered reachability policies. The main difference with respect to the synthetic case is that I3-I5 exhibit a considerably higher running time. This increase in running time can be attributed to the overall increased complexity of the considered realistic configurations. Indeed, the main contributor to this increase is the time spent to find the solution for the MaxSAT problem. Although, due to data confidentiality, we cannot provide details regarding the distribution of the number of required configuration changes, we can say that in some cases AutoNet was forced to apply 10+ different configuration changes to satisfy some reachability intents. Such scenarios reflect in \cref{fig:intents_cn5} with an increased variance for some of the results.

These results, combined with results from our scalability analysis, show that AutoNet is a highly promising approach for scalable and fast reachability policy enforcement for medium-to-large scale cloud deployments.

\section{Related Work}\label{sec:related}
Numerous works in the field of network verification~\cite{hh-analysis, zhang2020apkeep, yang2015real, prabhu2020plankton, anteater, netplumber, lopes2015checking, batfish, minesweeper} tackled the problem of scalability and correctness of data plane verification. For this reason, multiple techniques such as BDD~\cite{batfish}, set algebra~\cite{hh-analysis}, and SAT-based encodings~\cite{minesweeper}, have been employed. These techniques target the reduction of the complexity of the problem by efficiently summarizing complex header spaces and their interactions with the network services.
Nevertheless, few attempts have been made in the direction of \emph{incremental data plane configuration synthesis} as opposed to simple configuration verification. 

Perhaps the most relevant works to AutoNet are~\cite{bayless2021debugging}, and~\cite{saha2015netgen}. Both systems propose an SMT-based framework for debugging reachability property violations within a VPC network. The system in~\cite{bayless2021debugging} is capable of extracting the minimum configuration set that is required to be changed in order to enforce some desired connectivity property. Yet, while it can provide suggestions on how to fix a potential misconfiguration, it neglects any effect of the fix on the rest of the network. Thus, leading to the possibility of introducing security policy violations and/or service interruptions if applied blindly. \cite{saha2015netgen} follows a similar approach, yet it targets SDN data planes, thus not taking full advantage of the topological properties of VPC networks, while at the same time not supporting replicated and tangled configuration, and any complex data plane service beyond routers and firewalls.

The authors of Jinjing~\cite{tian2019safely} make use of SAT-based network encoding to reason about different ACL-related properties within a network. Notably, Jinjing is capable of synthesizing ACL configurations following a particular user specification. Nevertheless, the system is only capable of reasoning about ACLs and does not consider more complex network services such as routers, load balancers, NAT-GW, etc. Additionally, Jinjing can only change existing configurations as opposed to AutoNet that, when needed, can  generate entirely new network services within the cloud deployment. 

NetComplete~\cite{el2018netcomplete}, Propane/AT~\cite{beckett2017network}, NetEgg~\cite{yuan2014netegg}, AED~\cite{abhashkumar2020aed} are capable of synthesizing network configurations from scratch, yet they targets mainly control plane protocols.

Finally, in addition to individual limitations, none of the aforementioned techniques can take into account the user-specified cost function used by AutoNet over which to perform the optimization process.

\section{Discussion and future work}\label{sec:discussion}
While we have evaluated AutoNet on production data showing its versatility and scalability there are still challenges that remain unaddressed.

\textbf{Inter-vpc hopping:}
Perhaps the biggest limitation of AutoNet is the absence of support for inter-VPC hopping. Indeed, AutoNet is not capable of reasoning about traffic scenarios in which a flow would hop between multiple VPCs before landing on the destination VPC. This limitation will not only introduce false positives as in the previous two cases but may also affect the safety of the system. Nevertheless, this behavior does not affect the correctness of the reachability policy enforcement and only requires a slight change to the SAT formulation to restore safety. This point, alongside the previously discussed points related to the SAT formulation, is an active part of our ongoing work.

\textbf{Service coverage:}
Additionally, AutoNet is not complete in terms of the supported network services. Notably, we did not include hybrid cloud network services such as Enterprise Switch and Enterprise Router
 in AutoNet. This choice was made due to their novelty at the time of the development of AutoNet and not due to the technological limitations adopted by AutoNet. Indeed, both services can be easily reduced down to a common abstraction used by AutoNet and encoded within the formulation. We plan to support them in the next releases of AutoNet in the near future.

\textbf{Comprehensive objective definition engine:}
Throughout the paper, we discussed the user-defined objective weights to be used within the MaxSAT solution. While being sound from the theoretical point of view, from the practical point of view a customer-facing technology that requires the definition of such many parameters may be burdensome and confusing. In the future, we plan to provide an automatic intent-based weight assignment engine capable of synthesizing a suitable set of weights from high-level intents (e.g., minimizing the overall cost of the deployment, minimizing the configuration complexity, etc.). A suitable abstraction level for this subsystem is still under discussion and is planned for future releases.

\section{Conclusion}\label{sec:conclusion}
In this paper, we introduce AutoNet, a system for automatic network reachability management in cloud networks. AutoNet provides a comprehensive abstraction that hides all of the complexity of managed cloud network services behind an simple abstraction while achieving sub-second response time. At its core, AutoNet uses network pre-filtering and quantization to reduce the complexity of the reachability enforcement problem. At the same time, it uses SAT-based encoding of the network combined with topological properties of the network to achieve soundness.
Our experiments show that, even for large VPC deployments, AutoNet is able to enforce the desired reachability policies in under 10 seconds, while at the same time, for more realistic intents the running time of AutoNet stays below 250ms. We argue that AutoNet can be seen as a full-fledged alternative to traditional cloud network abstractions and represents a step further towards fully autonomous cloud networks operation.

\bibliographystyle{IEEEtran}
\bibliography{biblio}

\end{document}